\pdfoutput=1
\documentclass[useAMS,usenatbib]{mn2e}
\usepackage{graphicx,subfig,lscape}
\usepackage[usenames,dvipsnames]{xcolor}

\newcommand{\vs}{$v \sin i$}
\newcommand{\teff}{$T_{\rm eff}$}
\newcommand{\lgg}{$\log\,{g}$}

\title[MDI of Ap Stars III: Abundance mapping of $\alpha^2$ CVn]{Stokes $IQUV$ magnetic Doppler imaging of Ap stars - III. Next generation chemical abundance mapping of $\alpha^2$ CVn\thanks{
Based on observations obtained at the Canada-France-Hawaii Telescope (CFHT) which is operated by the National Research Council of Canada, 
the Institut National des Sciences de l'Univers of the Centre National de la Recherche Scientifique of France,  and the University of Hawaii.  Also based on observations obtained at the Bernard Lyot Telescope (TBL, Pic du Midi, France) of the Midi-Pyr\'en\'ees Observatory,  which is operated by the Institut National des Sciences de l'Univers of the Centre National de la Recherche Scientifique of France. }
}

\author[J.Silvester, O.Kochukhov, G.A. Wade]
{J. Silvester$^{1,2}$, O.Kochukhov$^{3}$, G.A. Wade$^{2}$ \\
$^{1}$Department of Physics, Engineering Physics \& Astronomy, Queen's University, Kingston, Ontario, Canada, K7L 3N6\\
$^{2}$Department of Physics, Royal Military College of Canada, P.O. Box 17000, Station `Forces', Kingston, Ontario, Canada, K7K 7B4\\
$^{3}$Department of Astronomy and Space Physics, Uppsala University, 751 20, Uppsala, Sweden\\
}  
\begin{document}

\date{Accepted . Received }

\pagerange{\pageref{firstpage}--\pageref{lastpage}} \pubyear{2014}

\maketitle

\label{firstpage}

\begin{abstract}
In a previous paper we presented an updated magnetic field map for the chemically peculiar star  $\alpha^2$~CVn using ESPaDOnS and Narval time-resolved high-resolution Stokes $IQUV$ spectra. In this paper we focus on mapping various chemical element distributions on the surface of $\alpha^2$~CVn.  With the new magnetic field map and new chemical abundance distributions we can investigate the interplay between the chemical abundance structures and the magnetic field topology on the surface of $\alpha^2$~CVn. 

Previous attempts at chemical abundance mapping of  $\alpha^2$~CVn relied on lower resolution data. With our high resolution (R=65,000) dataset we present nine chemical abundance maps for the elements O, Si, Cl, Ti, Cr, Fe, Pr, Nd and Eu. We also derive an updated magnetic field map from Fe and Cr lines in Stokes $IQUV$ and O and Cl in Stokes $IV$. These new maps are inferred from line profiles in Stokes $IV$ using the magnetic Doppler imaging code {\sc invers}10.  We examine these new chemical maps and investigate correlations with the magnetic topology of  $\alpha^2$~CVn.  We show that chemical abundance distributions vary between elements,  with two distinct groups of elements; one accumulates close to the negative part of the radial field, whilst the other group shows higher abundances located where the radial magnetic field is on the order of 2 kG regardless of the polarity of the radial field component.  We compare our results with previous works which have mapped chemical abundance structures of Ap stars. With the exception of Cr and Fe, we find no clear trend between what we reconstruct and other mapping results. We also find a lack of agreement with theoretical predictions. This suggests that there is a gap in our theoretical understanding of the formation of horizontal chemical abundance structures and the connection to the magnetic field in Ap stars. 
 \end{abstract}

\begin{keywords}
stars: chemically peculiar - stars: magnetic field.
\end{keywords}

\section{Introduction}
The bright Ap star $\alpha^2$~CVn is a member of the class of chemically peculiar stars which also exhibit a strong globally ordered magnetic field. These chemical peculiarities are exhibited as global over and underabundances relative to the sun and as lateral abundance nonuniformities that have been described in the literature as spots or rings of over/under abundance, but can also be more complex (e.g  L\"{u}ftinger et al. 2003, Kochukhov et al. 2004 and Rice et al. 2004). Abundances are also reported to vary vertically through the atmosphere in a significant way. These abundance anomalies are believed to result principally from atomic diffusion in the atmosphere of the star (Michaud 1970). Michaud showed that if the atmosphere of a star is sufficiently stable, diffusion under the competing influence of gravity and radiation pressure can occur. The additional presence of the magnetic field strongly influences energy and mass transport (e.g., diffusion, convection and weak stellar winds) within the atmosphere of a star, and results in strong chemical abundance non-uniformities in photospheric layers (e.g., Turcotte 2003). Magnetic fields have been shown to modify diffusion in two ways. First, charged particles are strongly constrained to follow field lines. This can result in the magnetic field modifying the diffusion velocity (Alecian \& Stift 2006). Secondly, radiative accelerations are also modified by Zeeman desaturation (magnetically induced spectral line desaturation) and splitting of absorption lines (Alecian \& Stift 2006). 

Even with a theoretical framework for the formation of abundance anomalies,  very few observational studies have been made of Ap stars using chemical abundance mapping combined with magnetic field topology analysis from the same data (from medium to high resolution observations).  With the exception of previous work on $\alpha^2$~CVn (the subject of the present paper), notable examples include magnetic Doppler imaging of 53 Cam (Kochukhov et al. 2004) and mapping of the roAp star HD 24712  (L\"{u}ftinger et al. 2010). Examples of abundance Doppler imaging with a comparison to independent model of the magnetic field geometry include oxygen abundance structures mapped for the star $\theta$ Aur by Rice et al. (2004), multiple element mapping of $\varepsilon$ UMa by L\"{u}ftinger et al. (2003) and mapping of the roAp star HD 83368 by Kochukhov et al. (2004). A more recent example of abundance mapping performed (without simultaneously deriving the magnetic field) is HD 3980 (Nesvacil et al. 2012). In these studies, with the exception of a handful of cases, no clear correlations could be found between the magnetic field topology and the horizontal structures of most chemical elements. This was interpreted as a lack of up to date theoretical models predicting the formation of horizontal abundance structures. With such a small sample size there is still limited understanding of the interplay between specific chemical species in the photosphere and the magnetic field of Ap stars.  

A new magnetic map of the bright Ap star $\alpha^2$~CVn was reconstructed by Silvester et al. (2014) using Stokes $IQUV$ observations obtained with ESPaDOnS and Narval spectropolarimeters described by Silvester et al. (2012).  We demonstrated that the magnetic topology we derived agreed with that of Kochukhov \& Wade (2010) which used data taken a decade earlier. Importantly, we showed that the a similar magnetic field topology could be obtained by mapping different atomic line sets, with the only differences seen between the meridional field components.  Mapping of the distributions of the surface chemical abundances of several elements for $\alpha^2$~CVn using Stokes $IV$ was performed by Kochukhov at al. (2002), allowing a comparison between the local field properties and local photospheric chemistry. These original maps were limited by the small wavelength coverage of the SOFIN spectrograph and a lack of a detailed model of the field topology, which generally cannot be derived from Stokes $IV$ alone (Kochukhov at al. 2002, Kochukhov and Wade 2010). With the new spectropolarimetric (Stokes $IQUV$) observations described by Silvester et al. (2012), we will now investigate the chemical abundance structures of $\alpha^2$~CVn at a level of detail not previously possible, in particular due the increased wavelength coverage. By mapping the chemical surface structures of  $\alpha^2$~CVn and comparing them to our updated magnetic map,  we hope to further our understanding of how different chemical species are affected by the characteristics of the magnetic field. 

The paper is organised as follows: Section 2 briefly describes the observations, Section 3 discusses the procedure for selecting lines suitable for chemical abundance mapping. In Section 4 we discuss the chemical abundance maps and the implications of these maps. Finally we summarise our findings in the conclusion. 

\section{Spectropolarimetric observations}
Observations of $\alpha^2$ CVn were obtained between 2006 and 2010, with both ESPaDOnS and Narval spectropolarimeters. The full details of the observations and reduction are reported by Silvester et al. (2012), along with the log of observations. For this study only Stokes $IV$ profiles were used for abundance mapping due to the fact that the linear polarisation signatures were generally too weak for most of the studied elements to be useful for mapping.

\begin{table}
\begin{center}
\caption{Fundamental parameters used/derived for the $\alpha^2$ CVn mapping.  {\bf References}: (1) Kochukhov et al. (2002), (2) Farnsworth (1932), (3) Kochukhov and Wade (2010).}
\begin{tabular}{ccc}
\hline
\hline  
Parameter & Value & Reference \\
\hline
\teff &  $11600 \pm 500$ K  & (1) \\
\lgg & $3.9 \pm 0.1$ & (1) \\ 
$P_{rot}$ & 5.46939 days & (2) \\
\vs & $18.0 \pm 0.5 $ km/s & \\
$i$ & $120^{\circ} \pm 5$ & (3) \\
$\Theta$ & $115^{\circ} \pm 5$ & (3) \\
\hline
\label{parameter-table}
\end{tabular}
\end{center}
\end{table}

\begin{table}
\begin{center}
\caption{Atomic lines used for  the $\alpha^2$ CVn mapping. The $\log gf$ values are those provided by the Vienna Atomic Line Database (VALD, Kupka et al. 1999), the primary references are given where possible and are as follows: (1) NIST10 (Ralchenk et al. 2010), (2) Schulz-Gulde (1969), (3) Blanco et al. (1995), (4) Matheron et al. (2001), (5) Kurucz (2012), (6) Wiese et al. (1969), (7) Wood et al. (2013), (8) Raassen \& Uylings (1998), (9) Ryabchikova (2006), (10) DREAM database Biemont et al. (1999), (11) Lawler et al. (2001).}
\begin{tabular}{cccc}
\hline
\hline  
Ion & Wavelength & $\log gf$ & Ref \\
&(\AA )  & \\
\hline
O~{\sc i} & 7771.941 &  0.369 & (1)\\
       & 7775.388 & 0.001 & (1) \\
\hline         
 Si~{\sc ii}  & 5055.984 & 0.593 & (2,3,4) \\
          & 5056.317 & -0.359 & (2,3) \\
          & 5978.930 & 0.040 & (2,3) \\
          & 6347.109 & 0.297 & (2,3,4) \\
          & 6347.133 & -1.200 & (5) \\
          & 6347.197 & -2.350 & (5) \\
          & 6371.371 & -0.040 & (2,3,4) \\
\hline         
 Cl~{\sc ii}  & 4794.556 & 0.455 & (6) \\   
          & 4819.480 & 0.064 & (6) \\
          & 4819.756 &  -0.790 & (6) \\
          & 4904.776 & 0.310 & (6) \\
\hline
 Ti~{\sc ii}  & 4163.644 & -0.130 & (7) \\
          & 4468.507 & -0.600 & (7) \\
          & 4571.960  &  -0.320 & (7)  \\
\hline
Cr ~{\sc ii} &  4588.199  & -0.845 & (8) \\
         & 4592.049 & -1.473  & (8) \\
         & 5246.768 & -2.560 & (8)  \\
         & 5279.876 & -2.112 & (8)  \\
         & 5280.054 & -2.316  & (8)  \\
\hline
Fe ~{\sc ii}& 4555.893 & -2.421 & (8) \\
          &  5030.630 &  0.431 & (8) \\
          &  5032.712 &  0.077 & (8) \\
          & 5035.708 & 0.632  & (8)\\ 
\hline
 Nd ~{\sc iii} & 4927.420 & -0.800 & (9)\\
             & 5050.695 & -1.060 & (9) \\
             & 5677.120  &-1.450 & (9) \\
             & 6145.068 & -1.330 & (9) \\
 \hline
 Pr ~{\sc iii}& 5299.993 &  -0.530 & (10) \\
            & 5765.243 & -1.100 & (10) \\
            & 7030.390 & -0.780 & (10) \\
  \hline
Eu ~{\sc ii}  & 6437.640 & -0.320 & (11) \\
           & 6645.100 &  0.120 & (11) \\
\hline
\label{line-list}
\end{tabular}
\end{center}
\end{table} 

\begin{figure*}
\begin{center}
       \includegraphics[width=0.85\textwidth]{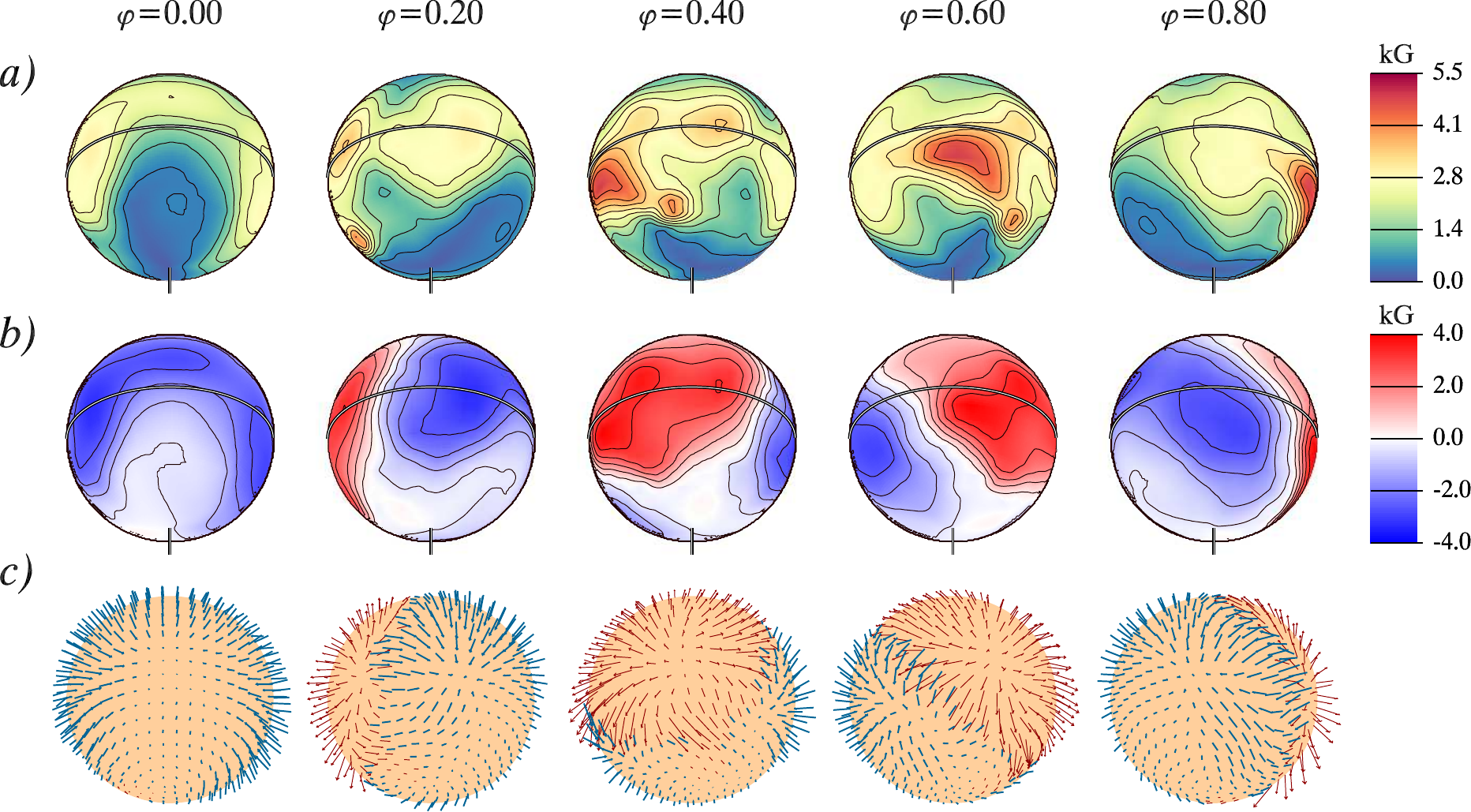}
       \includegraphics[width=0.85\textwidth]{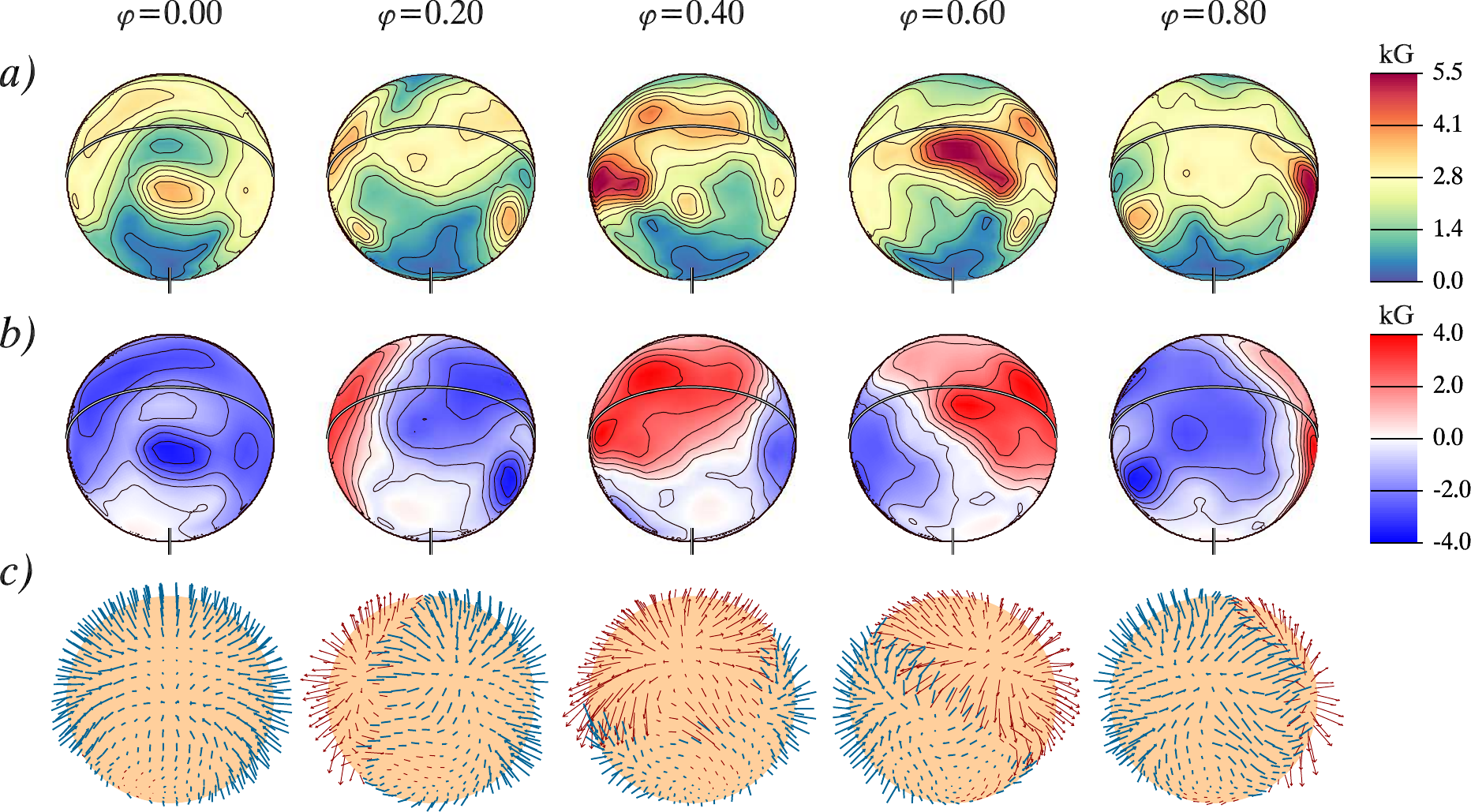}
     \caption{Magnetic maps of $\alpha^2$ CVn (top) computed using {\sc Invers10} for all selected chromium lines and iron lines from Silvester et al. (2014)  and (bottom) the new updated magnetic field map for which we included the same lines as Silvester et al. (2014), plus the addition of oxygen and chlorine lines in Stokes $IV$. The spherical plots show distributions of the field modulus (a), radial field (b) and field orientation (c) and each column is a different phase of rotation (0.0, 0.2, 0.4, 0.6 and 0.8).}
\label{Maps-field}
\end{center}
\end{figure*}

\begin{figure*}
\begin{center}
 \includegraphics[width=0.85\textwidth]{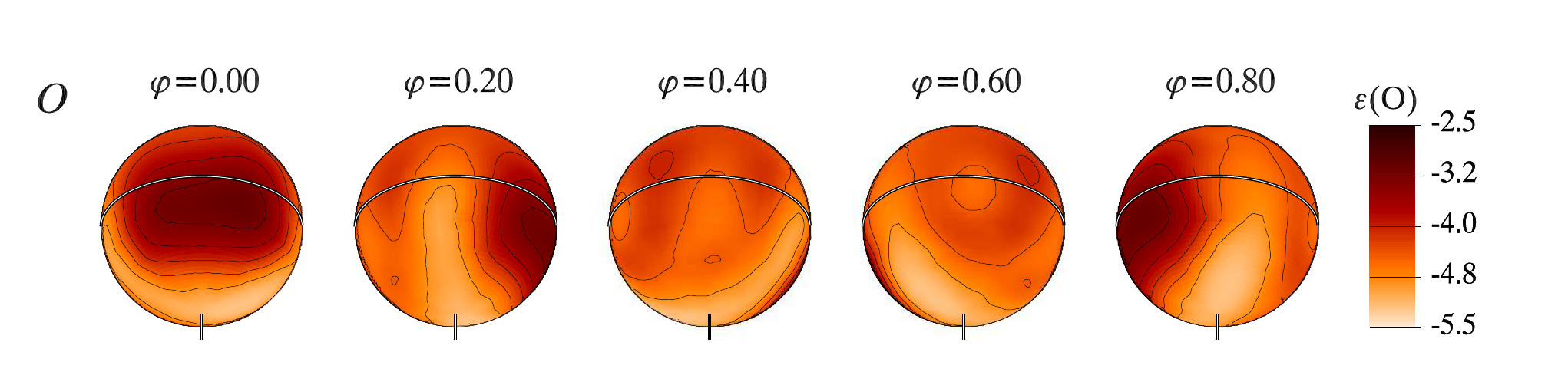}
  \includegraphics[width=0.85\textwidth]{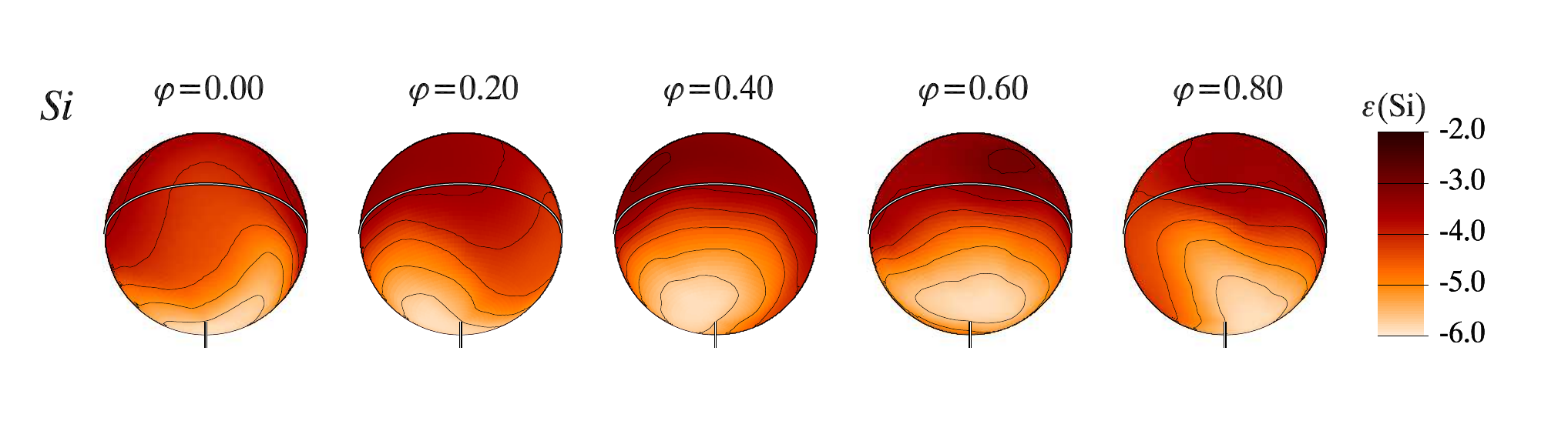}
 \includegraphics[width=0.85\textwidth]{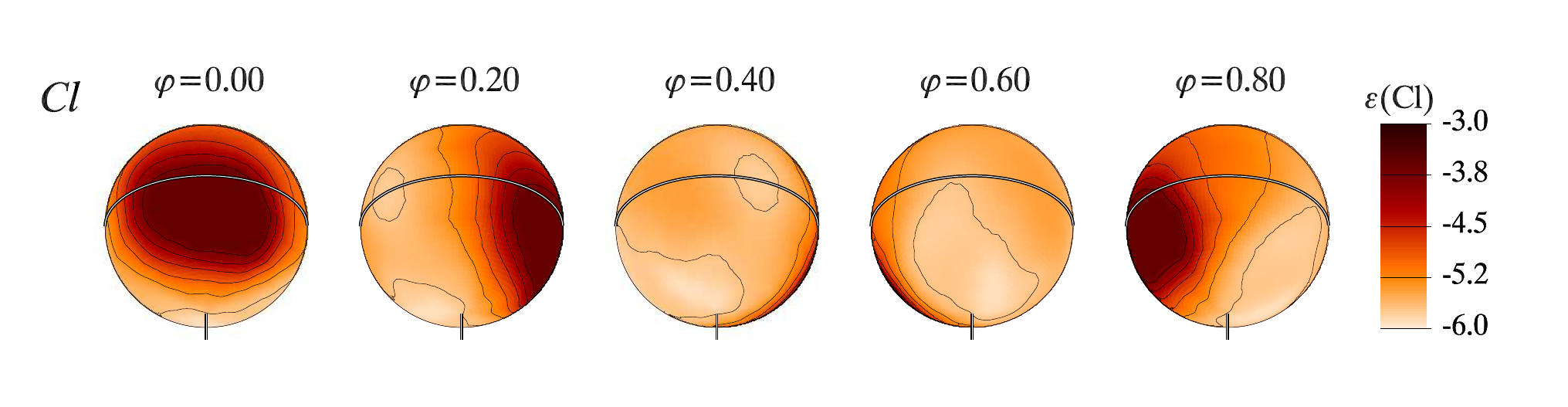}
  \includegraphics[width=0.85\textwidth]{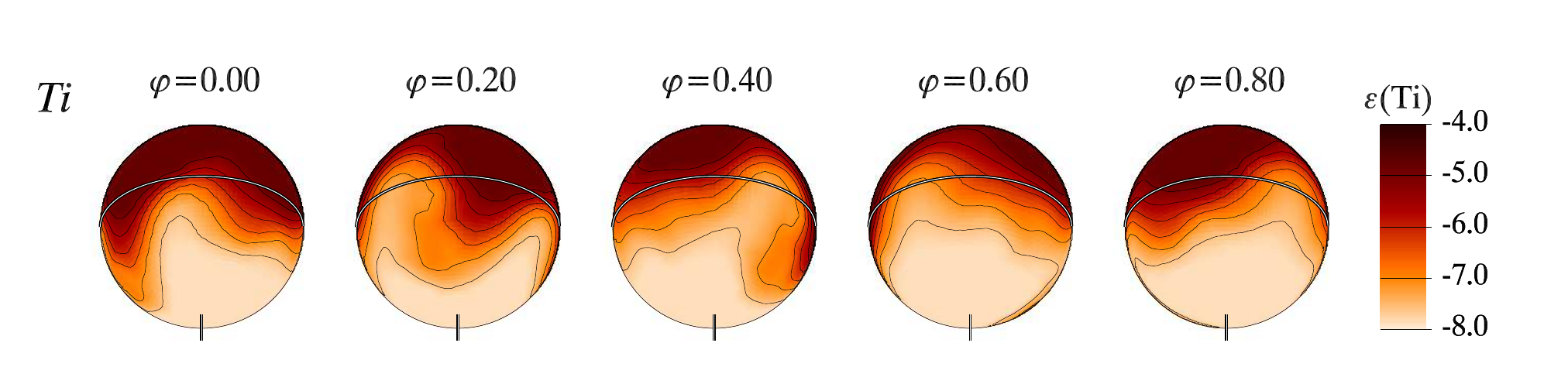}
  \includegraphics[width=0.85\textwidth]{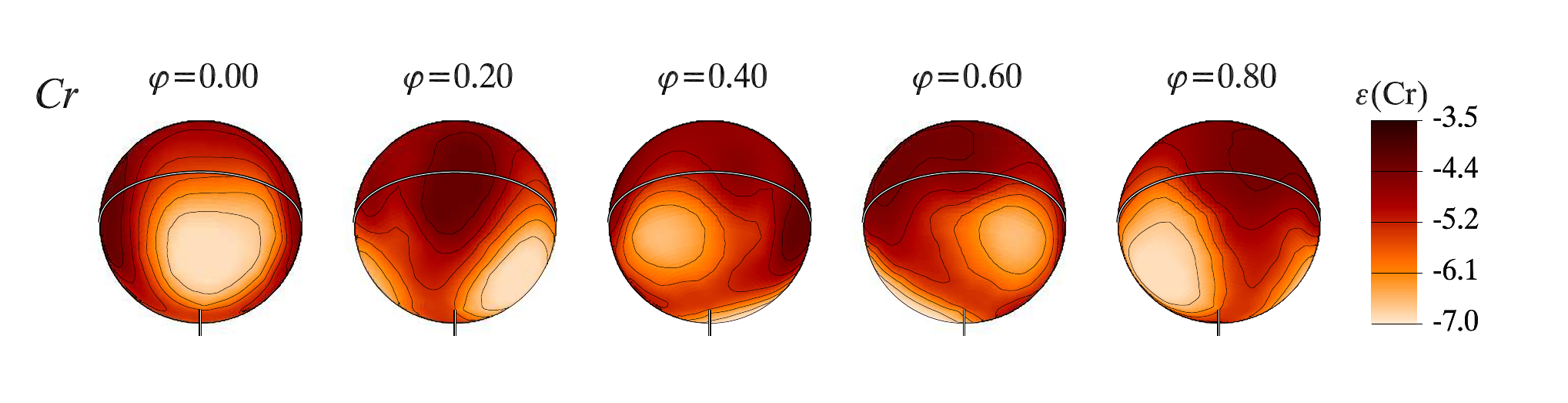}
  \includegraphics[width=0.85\textwidth]{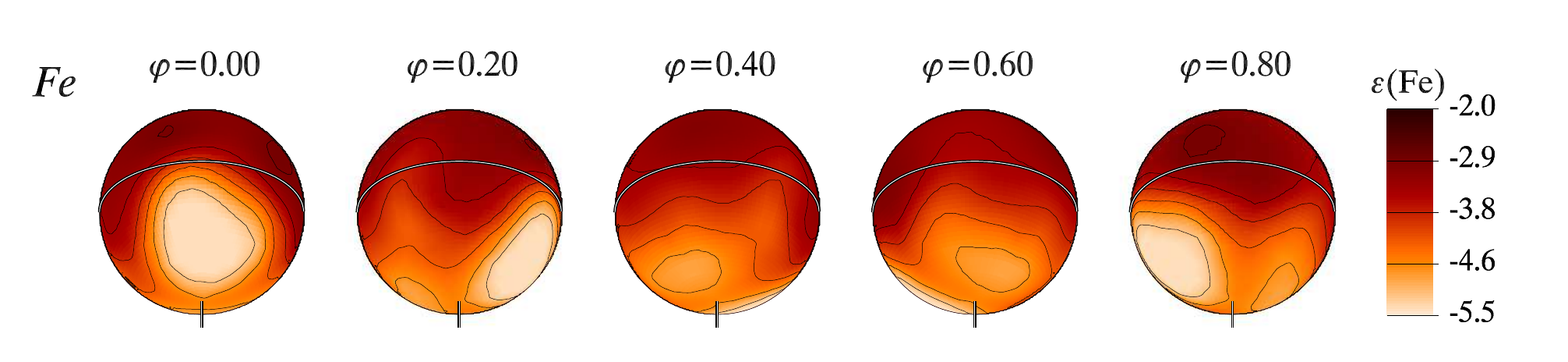}
     \caption{Chemical abundance distributions based on the observations of Silvester et al. (2012) using {\sc Invers10} for light and iron peak elements: O, Si, Cl, Ti, Cr, and Fe. Each column indicates a different rotational phase (0.0, 0.2, 0.4, 0.6 and 0.8) and the solid line shows the location of the stellar equator. The visible rotational pole is also indicated.}
\label{Maps1}
\end{center}
\end{figure*}

\begin{figure*}
\begin{center}
    \includegraphics[width=0.85\textwidth]{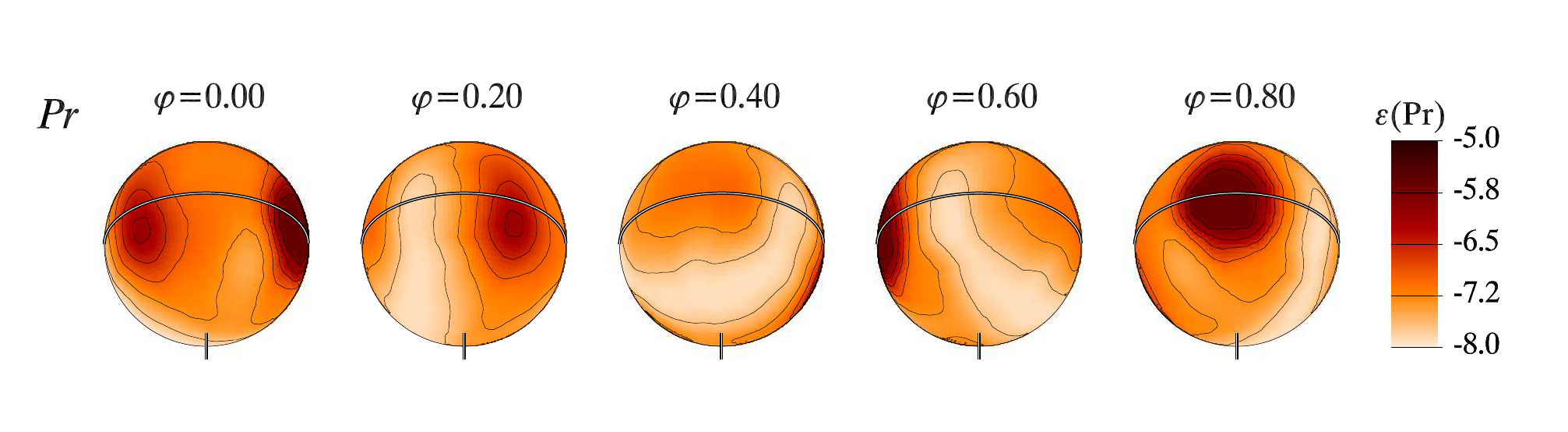}
    \includegraphics[width=0.85\textwidth]{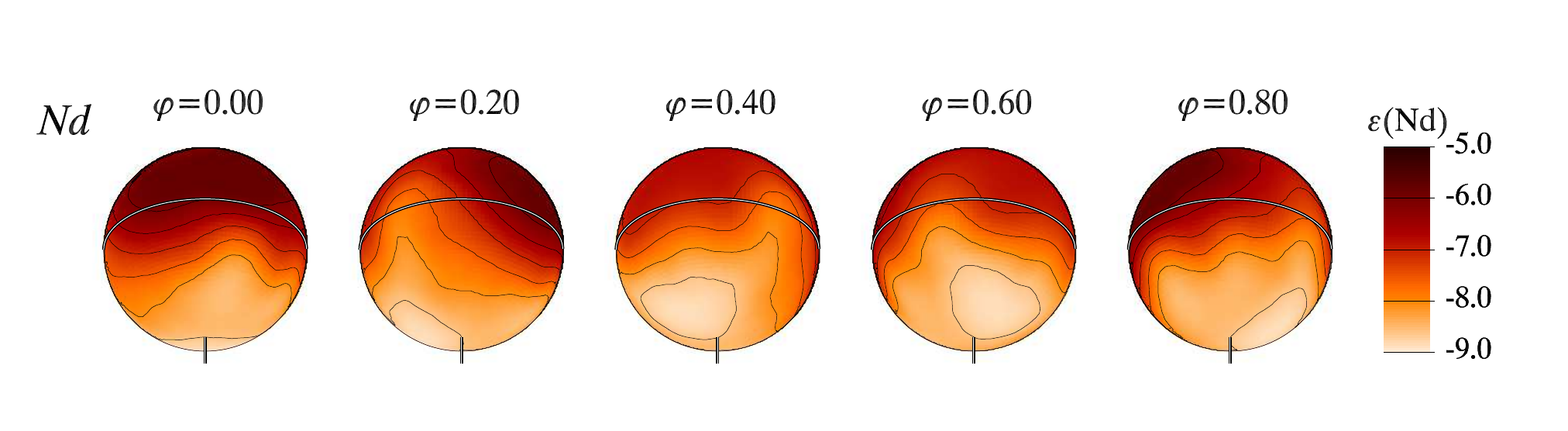}
    \includegraphics[width=0.85\textwidth]{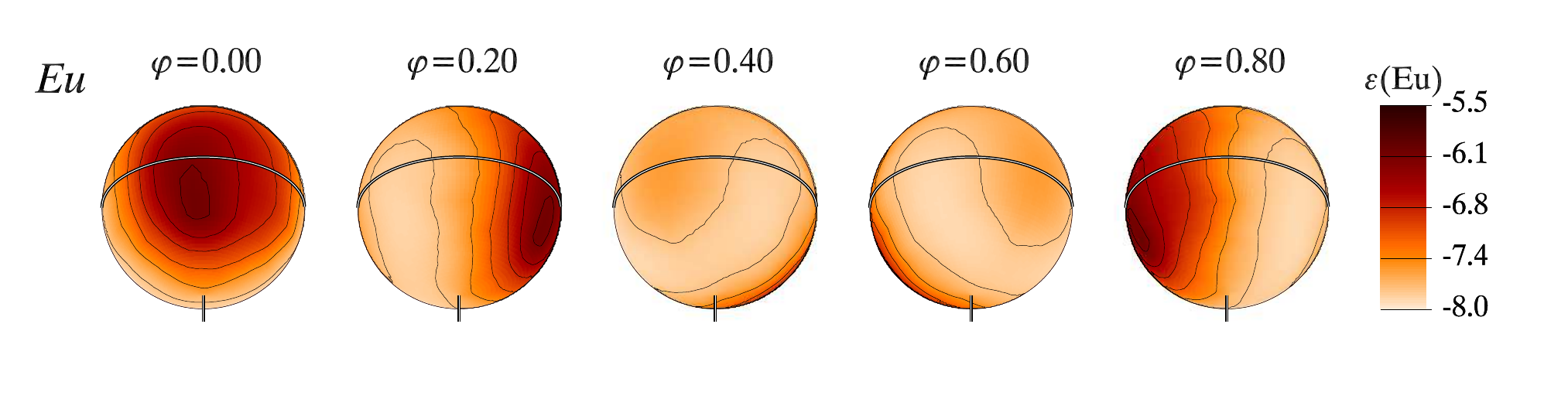}
     \caption{Chemical abundance distributions based on the observations of Silvester et al. (2012) using {\sc Invers10} for rare earth elements: Pr, Nd and Eu. Each column indicates a different rotational phase (0.0, 0.2, 0.4, 0.6 and 0.8) and the solid line shows the location of the stellar equator. The visible rotational pole is also indicated.}
\label{Maps2}
\end{center}
\end{figure*}

\section{Chemical and Magnetic Map Inversion}
The methodology used to derive magnetic field maps is described by Silvester et al. (2014).  In this paper we will concentrate on the abundance mapping of $\alpha^2$~CVn. The abundance mapping was performed using the {\sc invers}10 magnetic Doppler imaging (MDI) code (Piskunov \& Kochukhov 2002; and Kochukhov \& Piskunov 2002). As was described by Silvester et al. (2014), {\sc invers}10 is a stellar surface mapping code written in {\sc fortran} that constructs model line profiles based on an assumed initial spherical surface distribution of free parameters (in this case the element abundance) and then iteratively adjusts the parameters until the computed line profiles are in agreement with the observations. The code is parallelized and was run on an 8 CPU Mac Pro. Graphical output and file processing was performed using {\sc idl}. The time required for the code to converge to a solution for a set of $IV$ Stokes data (containing 28 phases), with fitting to one or two spectral lines, is of the order of an hour.

As described by Silvester et al. (2014) the fundamental parameters needed for inversion have to be well defined.  Table \ref{parameter-table} shows the key parameters used in this study. The basis of line selection for abundance mapping combined a visual inspection of the spectra looking for candidate lines which were unblended and also a review of the literature looking for unblended lines that have been used in previous studies for stars of a similar spectral type. The selection drew on the line lists of of Pyper (1969), Cohen (1970) and to a lesser extent Roby and Lambert (1990) and those used by Bailey et al. (2014).  The next stage was then to eliminate lines from the selection list which did not show strong line depths.  Because in this paper we only focus on chemical abundance maps, only Stokes $IV$ profiles were used in the mapping and unlike the requirement by Silvester et al. (2014) candidate lines did not have to exhibit linear polarisation signatures. To ensure that the resulting abundance maps were of sufficient reliability, only elements which had a minimum of two lines suitable for inversion were considered. 

An important parameter in the reconstruction of the abundance maps is the choice of the regularisation. As is described by Piskunov \& Kochukhov (2002) {\sc invers}10 uses a Tikhonov regularisation function, which assists the code in converging to a solution by providing a limit on how smooth or patchy the resulting map can be. For all the maps a value of regularisation was chosen which gave the lowest total discrepancy, whilst still reproducing the Stokes profiles without fitting to a significant amount of noise. However a lower limit was set such that the total regularisation scaling factor must be at the minimum only a factor of 10 times smaller than the total discrepancy (between the observations and the model). After this limit the improvement to the discrepancy becomes increasingly smaller, but the map will continue to become increasingly patchy.  

\begin{figure}
\begin{center}
       \includegraphics[width=0.42\textwidth]{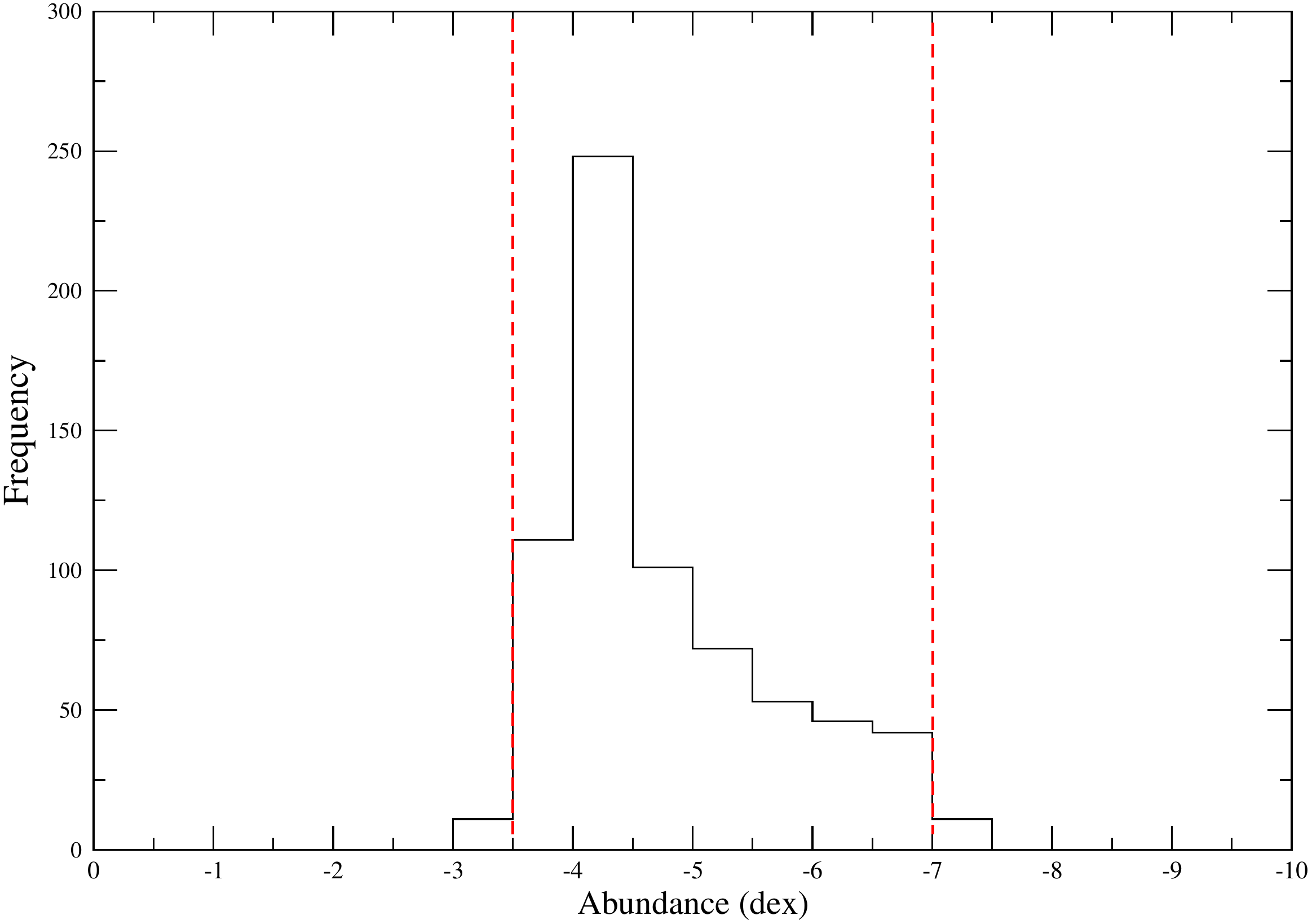}
     \caption{Histogram showing the range of abundance values found in the reconstructed map for chromium. The red dashed lines indicate the abundance range cut off, after which values are not included in the abundance range scale. In this case an abundance range of -3.5 to -7.0 dex was used as the scale for chromium.}
\label{histo}
\end{center}
\end{figure}

When producing the final spherical plots for each derived chemical map, the choice of abundance scale range has to be taken with care. In particular it is best to avoid using an abundance scale with an extremely large range if this range is based on a handful of outlying abundance values, otherwise the resulting map may overemphasise unrealistic abundance features. {\sc invers}10 outputs map data in the form of an ascii file which contains a numerical abundance value for each of the surface elements (in this case 695) which make up the complete spherical map. These 695 abundance values were plotted in histogram form to identify any extreme outliers,  allowing the abundance scale to be limited to a more realistic/representative range. An example histogram is shown for chromium in Fig. \ref{histo}, with red dashed lines indicating the abundance range cut off. This process was performed for all the chemical abundance maps presented and in some cases on the order of $\approx$20 points were ignored. In addition because of the inclination of the star,  abundance values at latitudes higher than +60$^{\circ}$ are harder to constrain, so abundance values corresponding to these latitudes were given less weight in the scale selection process. In fact in most cases these values contributed considerably to the outlier values and thus were typically already discarded from the abundance range selection. 

\section{Refined Magnetic Field Map}
For all of the abundance map inversions the magnetic field was set as a fixed parameter: the magnetic field reconstructed from both chromium and iron lines and referred to as the ``final map'' by Silvester et al. (2014) was initially used for this fixed magnetic field. But when independently mapping the abundances of oxygen and chlorine it was noticed that whilst the fit for Stokes $I$ was very good, the fit for Stokes $V$ was quite poor, with clear disagreement of the amplitude of Stokes $V$ at various phases between the model and the observations. 

To investigate if this poor fit in Stokes $V$ was due to oxygen and chlorine lines emphasising a different region of the magnetic field map to which iron and chromium lines are less sensitive, we recalculated the magnetic field map. We used the same Stokes $IQUV$ iron lines (Fe~{\sc ii}\ $\lambda$ 4923,  5018,  4273,  4520 and 4666) and chromium lines (Cr~{\sc ii}\ $\lambda$ 4824, 5246 and 5280) from Silvester et al. (2014), but also included Stokes $IV$ lines for oxygen (O~{\sc i}\ $\lambda$ 7771 and 7775) and chlorine (Cl~{\sc i}\ $\lambda$ 4794, 4819 and 4904). It should be noted that the Stokes $QU$ profiles for oxygen and chlorine showed no significant polarization detected and therefore were not included in the inversion. The resulting fit in Stokes $V$ for oxygen and chlorine was greatly improved with the updated magnetic map and at the same time the fit to Stokes $IQUV$ for iron and chromium was consistent with what was seen when using the magnetic map of Silvester et al. (2014). We therefore chose to adopt the updated magnetic map (Fe \& Cr Stokes $IQUV$ and O \& C Stokes $IV$) as the fixed magnetic map for all the inversions in this paper. Whilst the fits to Stokes $V$ were greatly improved for certain elements, the resulting abundance maps for these elements differed little from the maps reconstructed using the magnetic field map by Silvester et al. (2014).

The resulting map is shown with the ``final map'' of Silvester et al. (2014) in Fig. \ref{Maps-field}. This figure illustrates that by including Stokes $IV$ for oxygen and chlorine in the reconstruction, the resulting magnetic field is very similar to what is seen by Silvester et al. (2014), but with an additional small spot-like region seen at phase 0.00 below the stellar rotational equator. To retain the quality of the fit to Stokes $QU$ profiles required that the regularisation of the magnetic field to be reduced by a factor of 3 compared to what was used by Silvester et al. (2014). This reduction in regularisation should result in a slightly more ``patchy'' map, which is potentially the reason for the new structures seen at phase 0.00. This updated map does not change the findings of Silvester et al. (2014) as the new map is still broadly consistent with the ``final map'' by Silvester et al. (2014). In fact, at the phases where the small spot does not appear the field distributions are effectively identical. It should be noted that the magnetic mapping of Silvester et al. (2014) concentrated purely on lines which exhibited strong Stokes $QU$ signatures, which is why lines such as oxygen and chlorine were not included in this original magnetic mapping.

\section{Chemical Abundance Maps}

\subsection{Oxygen, Silicon and Chlorine Maps}
\begin{landscape}
\begin{figure}
\begin{center}
 \includegraphics[width=0.38\textwidth]{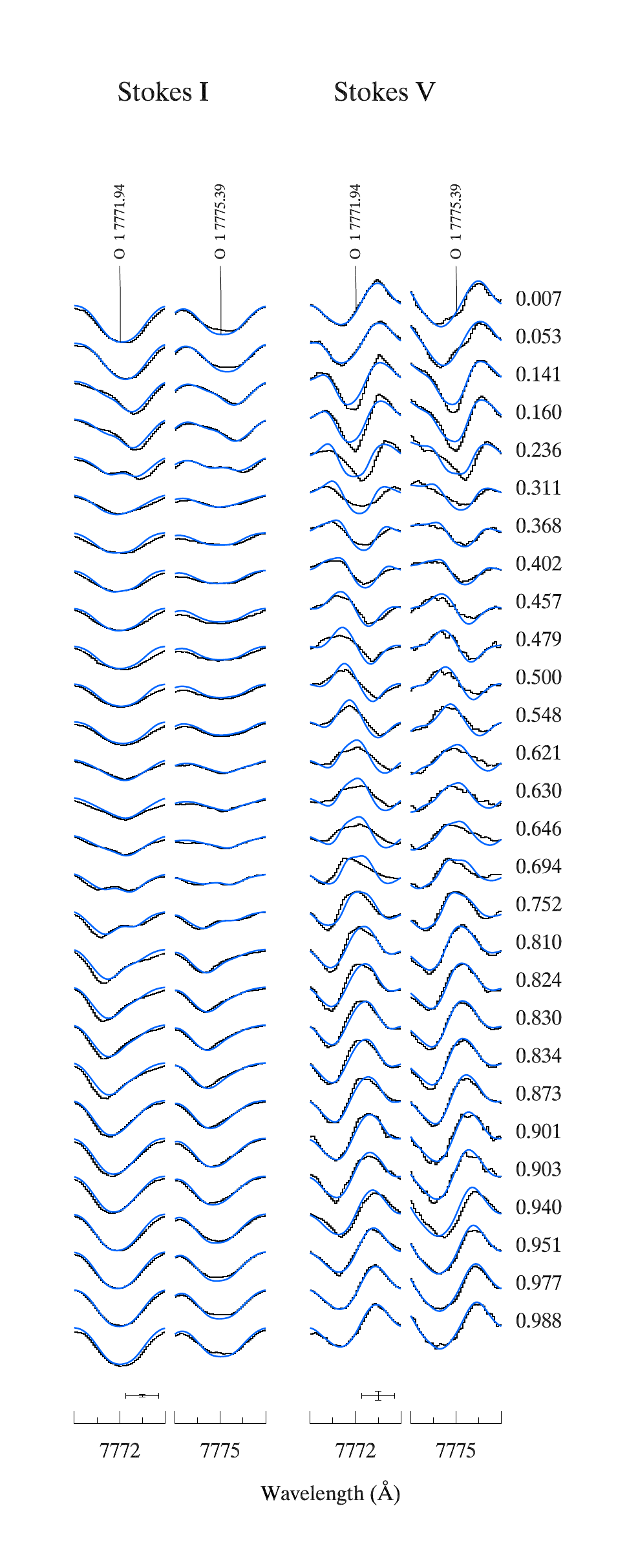}
    \includegraphics[width=0.52\textwidth]{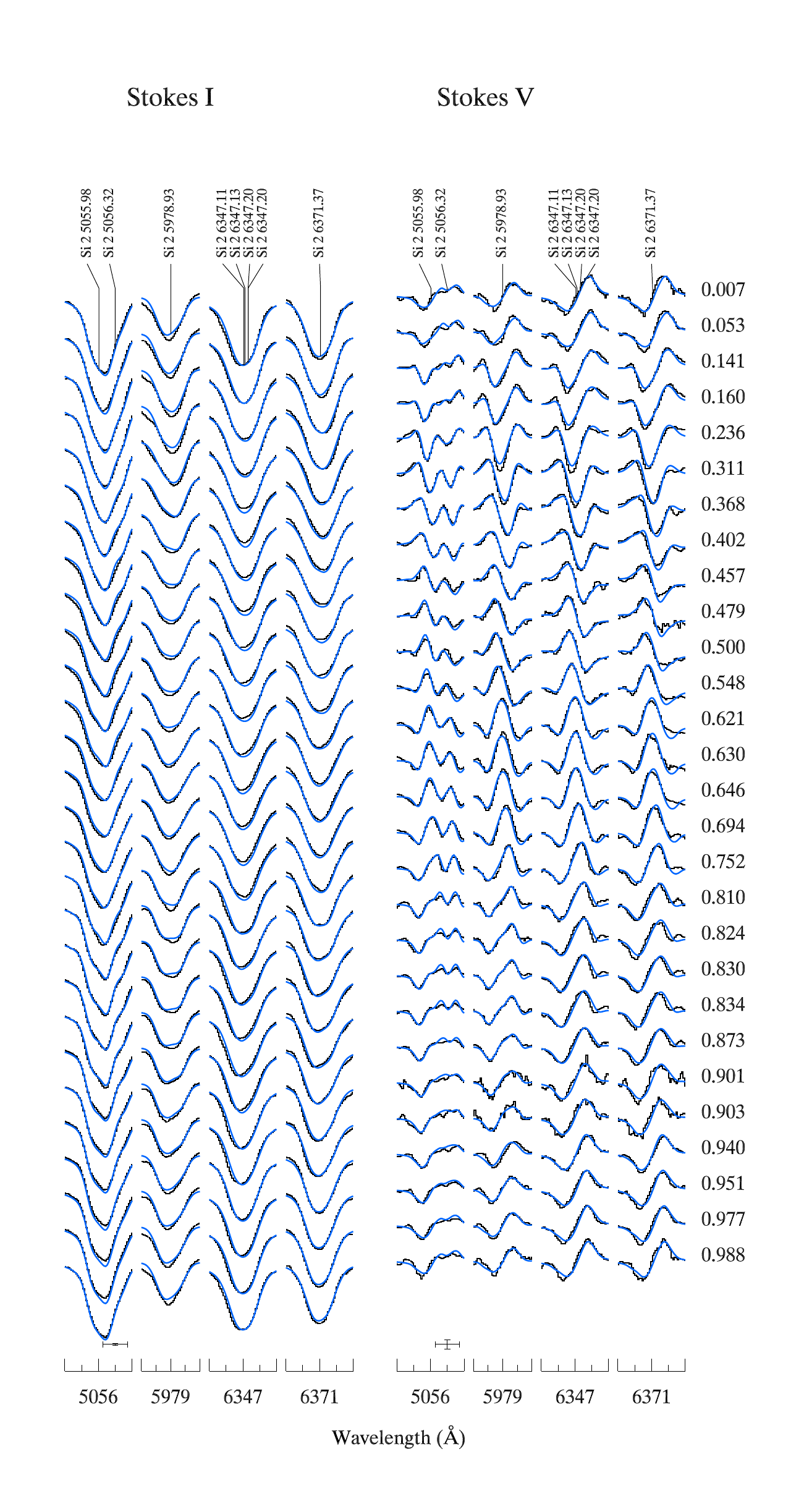}
   \includegraphics[width=0.40\textwidth]{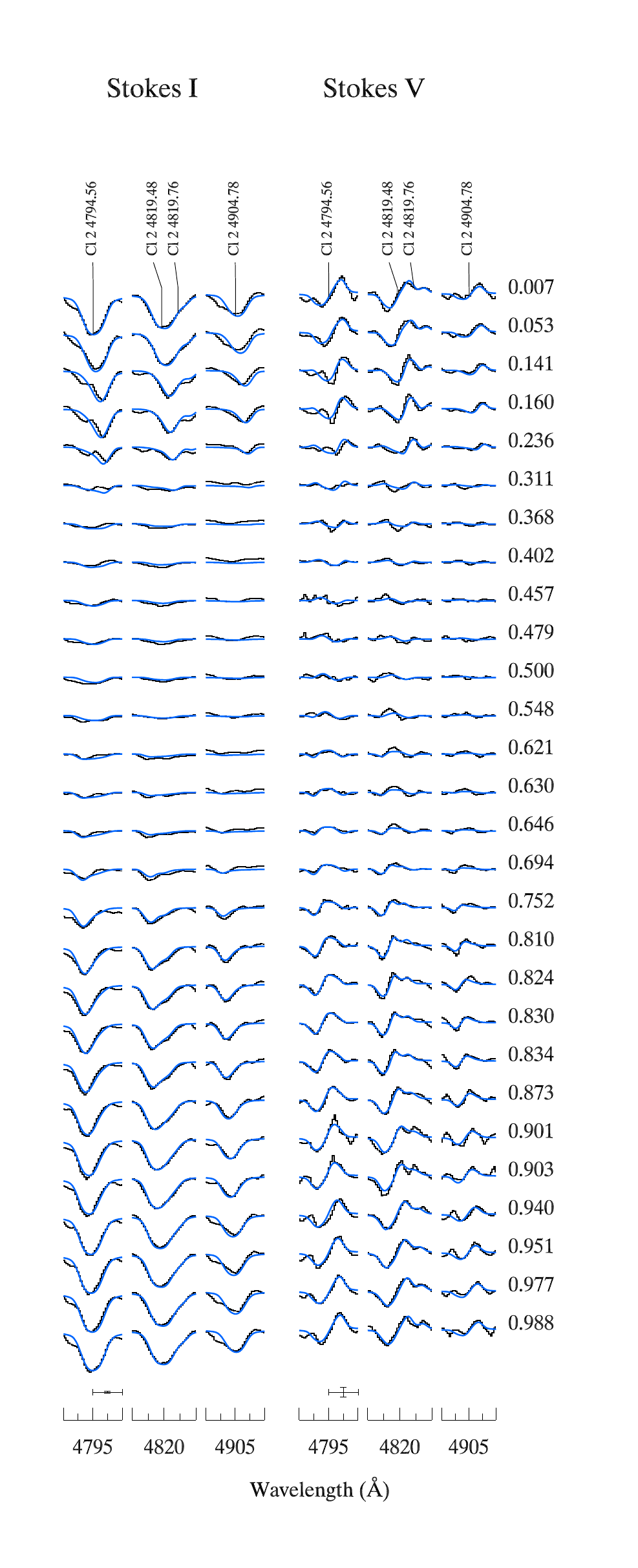}
   \caption{Comparison between observed (dots) and synthetic (solid curve,blue)  Stokes $IV$ parameter spectra of $\alpha^2$  using INVERS10. The solid curves shows the best fit to O, Si and Cl lines. }
\label{O-Mg-Si-fit}
\end{center}
\end{figure}
\end{landscape}

Oxygen chemical maps were produced using the Stokes $IV$ line profiles of O~{\sc i}\ $\lambda$ 7771 and 7775,  The fit between observations and the model can be seen in Fig. \ref{O-Mg-Si-fit} and the resulting map is shown in Fig. \ref{Maps1}. In Fig. \ref{O-Mg-Si-fit}, both lines show clear variability in the Stokes $I$ profile. The agreement between observations and the model is generally good for the Stokes $IV$, although there are some phases where the amplitude is not in agreement. This difference is likely due to non-LTE effects in these oxygen lines.  If this small discrepancy seen in Stokes $V$ is due to non-LTE effects and/or stratification, these effects can also influence the Stokes $I$ profile, however the code will still find good agreement in Stokes $I$ by adjusting the abundance accordingly, but such adjustments would have minimal effect on the Stokes $V$ fit. The end result of this is that whilst the abundance contrast will be realistic, the absolute abundance scale may not, but because it is the contrast that is of greatest importance, comparison with the magnetic field is still possible even if the absolute abundance is not fully realistic.

The oxygen abundance ranges from -2.5 dex to -5.5 dex (solar value = -3.21). A high abundance feature is clearly seen between phases of 0.80 to 0.20, comparing it to the magnetic topology from Silvester et al. (2014) (shown in Fig. \ref{Maps-field}),  the highest concentration of oxygen is found at latitudes between -45$^{\circ}$ to +45$^{\circ}$ and at longitudes between 300$^{\circ}$ and 60$^{\circ}$. This area aligns with the negative component of the radial magnetic field which is found at a similar latitude and longitude.   

Silicon chemical maps were produced using the Stokes $IV$ line profiles of Si~{\sc ii}\ $\lambda$ 5055, 5978,  6347 and 6371. The fit between observations and the model can be seen in Fig. \ref{O-Mg-Si-fit} and the map is shown in Fig. \ref{Maps1}. In Fig. \ref{O-Mg-Si-fit}, the lines show very slight variability in the Stokes $I$ profile and there is good agreement between the observations and the model in both Stokes $I$ and $V$.  Looking at the distribution of silicon, the abundance values vary from -2.0 dex  to -6.0 dex (solar value = -4.49),  the higher concentration areas are found at upper latitudes between -20$^{\circ}$ and 60$^{\circ}$ distributed over all longitudes. Comparing this to the magnetic map, we see the higher abundance areas correlate somewhat to the areas where the radial magnetic field (regardless of sign) has a field modulus of approximately 2 kG, close to the stellar equator between a range of latitudes of  -45$^{\circ}$ to +45$^{\circ}$. The lowest abundances are seen in locations where the magnetic field is weakest. 

A chlorine map was computed using line profiles of Cl~{\sc i}\ $\lambda$ 4794, 4819 and 4904. The fit between observations and the model can be seen in Fig. \ref{O-Mg-Si-fit} with strong variability in Stokes $I$ with the line profile almost disappearing at a phase of 0.3. There is good general agreement between the observations and model for Stokes $I$ and $V$. The reconstructed map is shown in Fig. \ref{Maps1}. Looking at the chlorine map,  the abundance ranges from -3.0 to -7.0 dex (solar value = -6.54), and somewhat similar to what was seen in oxygen, the high abundance structure is located at latitudes between -45$^{\circ}$ to +25$^{\circ}$ and longitudes between 320$^{\circ}$ and 40$^{\circ}$. Comparing this to the magnetic field we find a similar pattern to that seen for oxygen, with the higher concentration areas aligning with the negative part of the radial magnetic field.

\subsection{Iron Peak Elements - Titanium, Chromium and Iron Maps}
\begin{landscape}
\begin{figure}
  \includegraphics[width=0.43\textwidth]{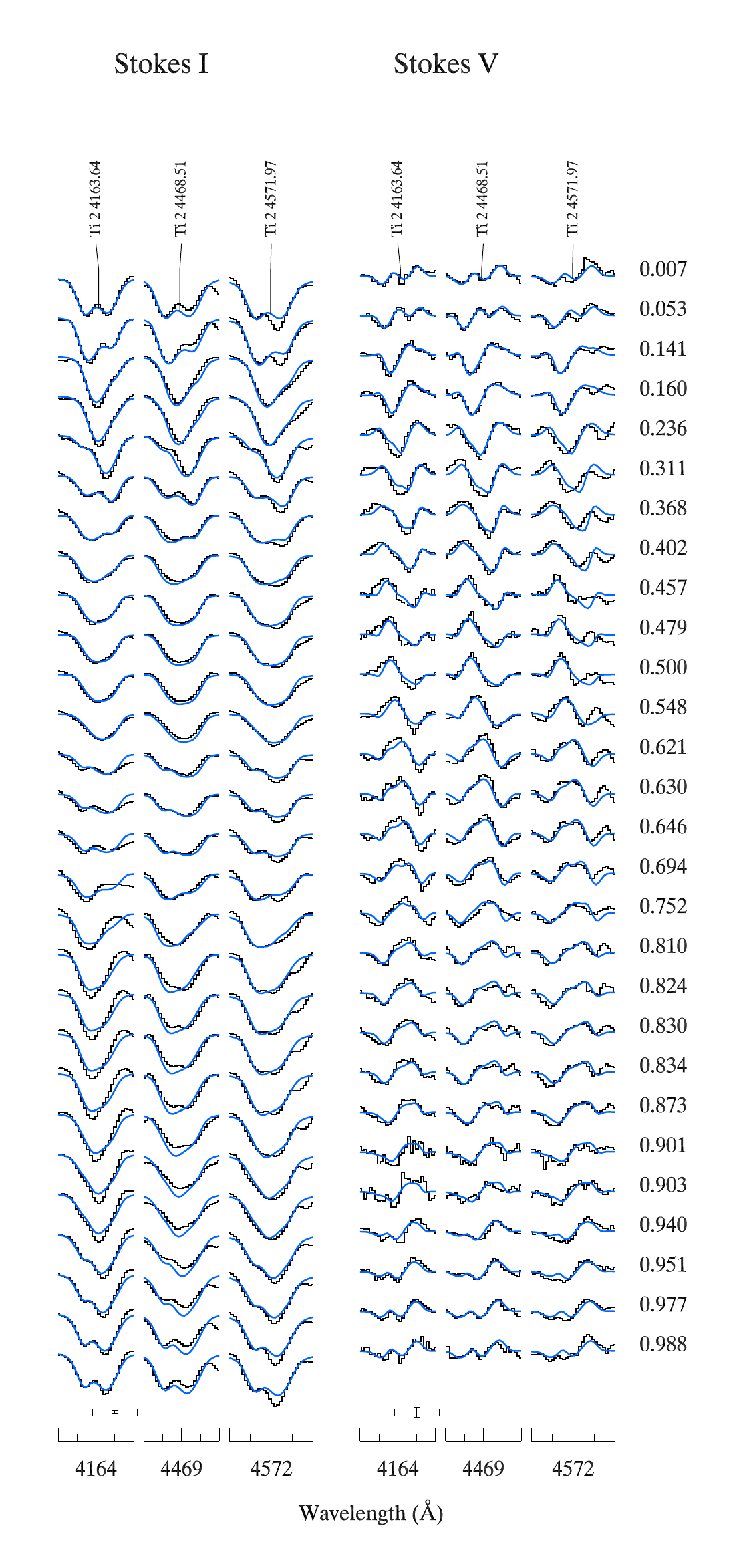}
 \includegraphics[width=0.44\textwidth]{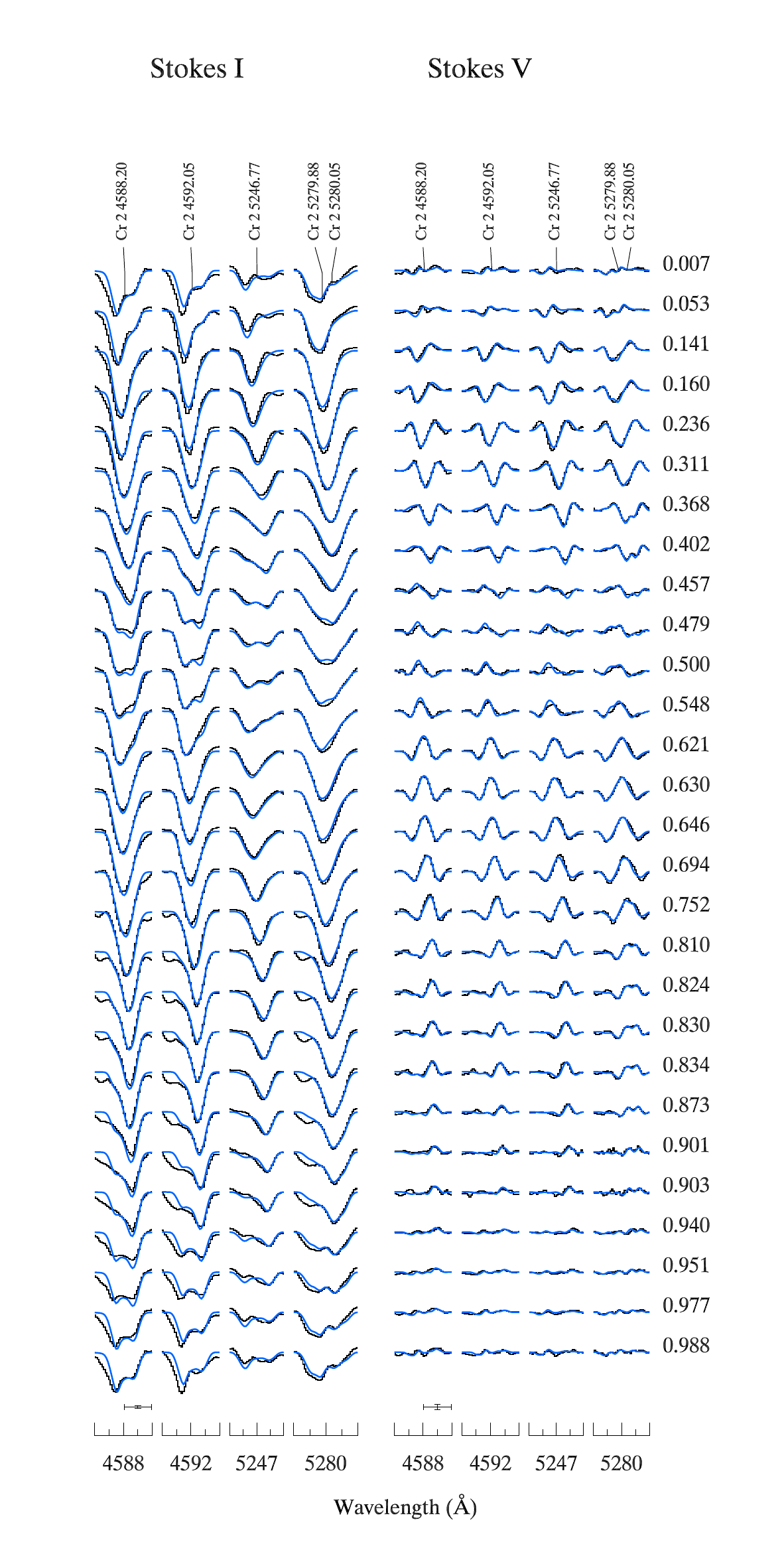}
\includegraphics[width=0.48\textwidth]{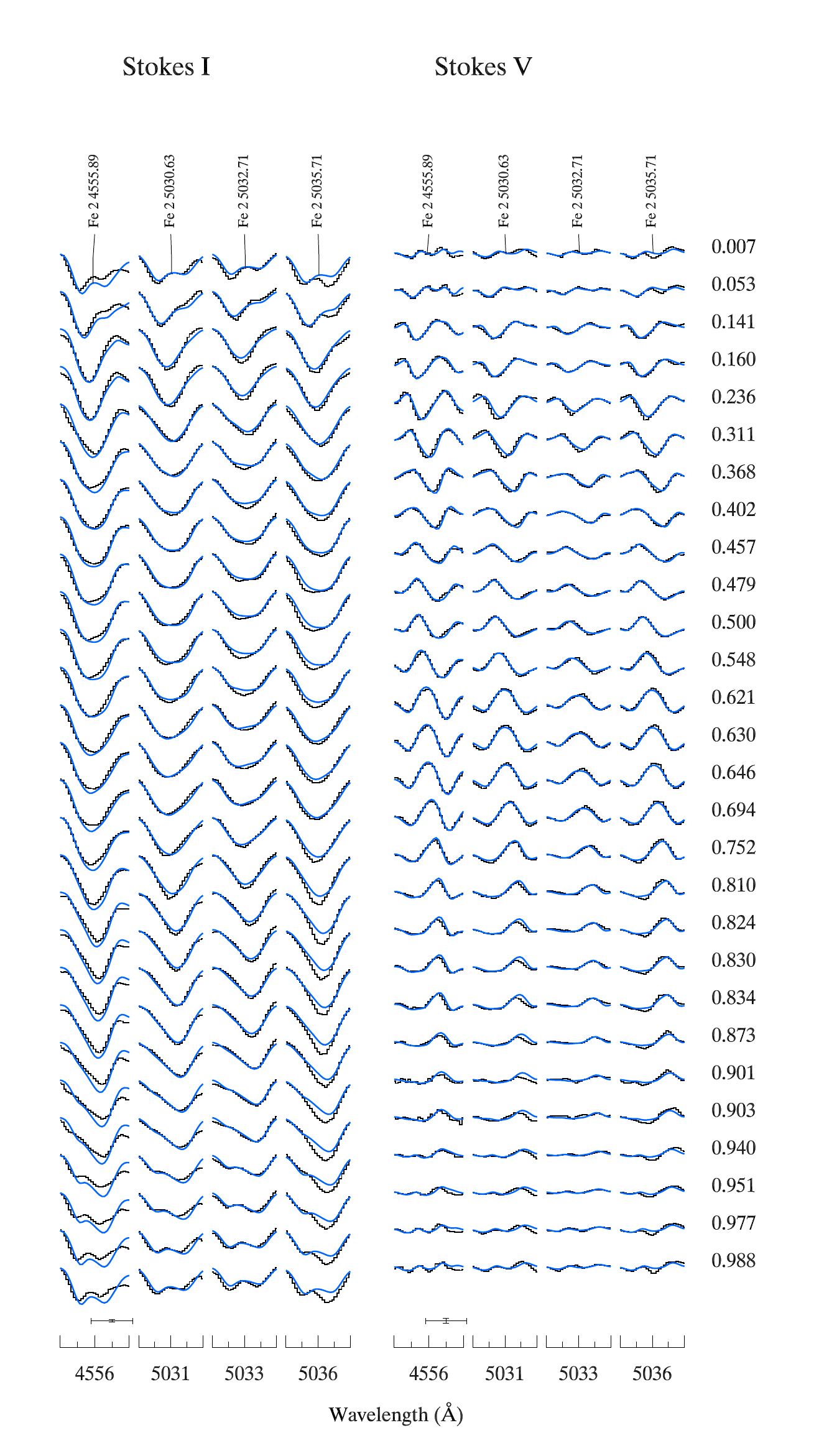}
   \caption{Comparison between observed (dots) and synthetic (solid curve,blue) Stokes $IV$ parameter spectra of $\alpha^2$  using INVERS10. The solid curves shows the best fit to Ti, Cr and Fe lines }
\label{Ti-Cr-Fe-fit-IV}
\end{figure}
\end{landscape}

Titanium chemical maps were produced using Stokes $IV$ line profiles of Ti~{\sc ii}\ $\lambda$ 4163, 4468 and  4571. The fit between observations and the model can be seen in Fig. \ref{Ti-Cr-Fe-fit-IV} and the map is shown in Fig. \ref{Maps1}.  In Fig. \ref{Ti-Cr-Fe-fit-IV}, all lines show very strong variability in the Stokes $I$ profile,  with a clear change in the line shape between phases. There is reasonable agreement between observations and the model,  but at some phases the wings in Stokes $I$ are not well fit.  Because we are fitting multiple line profiles simultaneously the final fit is a compromise between the different line profiles, therefore a perfect fit is not always expected. If a similar discrepancy is seen systemically in all lines, then this would be of concern. The titanium abundance ranges from -4.0 dex to -8.0 dex  (solar value = -7.02).  The abundance structure of titanium is somewhat similar to that of the silicon map, but with the higher abundance areas limited to latitudes of 0$^{\circ}$ to 60$^{\circ}$ and spread over all longitudes. When comparing this distribution with the magnetic field map, the higher abundance areas appear to correlate with areas where the radial field modulus is approximately 2 kG, regardless if it is the positive or negative component on the radial sphere which is seen between latitudes of -45$^{\circ}$ to 45$^{\circ}$. There are small discrete low abundance areas seen at longitudes 120$^{\circ}$ and 260$^{\circ}$ extending to just above the stellar equator. These are areas where the radial field is small in the transition region between negative and positive components. 

Chromium chemical maps using Stokes $IV$ were produced using the line profiles of Cr~{\sc ii}\ $\lambda$ 4588, 4592, 5246 and 5279. The fit between observations and the model can be seen in Fig. \ref{Ti-Cr-Fe-fit-IV} and the map is shown in Fig. \ref{Maps1}. In Fig. \ref{Ti-Cr-Fe-fit-IV}, all lines show very strong variability in the Stokes $I$ profile, with a clear change in the line shape between phases. Like the titanium fits, there is good agreement between observations and the model at most phases, some wings in Stokes $I$ are not well fit, but as described for titanium, this situation can arises when fitting multiple lines. The chromium abundance ranges from -3.5 dex to -7.0 dex  (solar value = -6.37).  The abundance structure is distinct from the light element abundance patterns,  with a very large low abundance feature located below the stellar equator, seen at latitudes -80$^{\circ}$ to -10$^{\circ}$ and at longitudes between 300$^{\circ}$ and 60$^{\circ}$. On the reverse side there is a slightly smaller low abundance structure seen at latitudes -60$^{\circ}$ to -10$^{\circ}$ and at longitudes between 140$^{\circ}$ and 220$^{\circ}$. The larger of these two structures appears to be located at an area where the radial magnetic field is negligible, this area is found at latitudes  -80$^{\circ}$ to -10$^{\circ}$ and a longitude between 300$^{\circ}$ and 60$^{\circ}$. Higher abundance areas are seen at latitudes -30$^{\circ}$ to 30$^{\circ}$ and at longitudes between 60$^{\circ}$ and 120$^{\circ}$ and 220$^{\circ}$ and 300$^{\circ}$. These trace the magnetic field in the areas where the radial field modulus is on the order of 2 kG. 

Iron chemical maps were constructed using the Stokes $IV$ line profiles of Fe~{\sc ii}\ $\lambda$ 4555, 5030 5032 and 5035. The fit between observations and the model can be seen in Fig. \ref{Ti-Cr-Fe-fit-IV} and the map is shown in Fig. \ref{Maps1}. In Fig. \ref{Ti-Cr-Fe-fit-IV}, all lines show variability in the Stokes $I$ profile, although the profile change is not as clear as with titanium and chromium. There is good agreement between observations and the model at most phases, there are phases where the core of the Stokes $I$ profiles are perfectly fit. As was described in the case of titanium, these small discrepancies can occur as a result of the compromises the inversion code has to make to fit all lines simultaneously. The iron abundance ranges from -2.0 dex to -5.5 dex  (solar value = -4.54), with the abundance structure being very similar to that found for chromium. This similarity extends to a large low abundance structure seen at latitudes -80$^{\circ}$ to -10$^{\circ}$ and at longitudes between 320$^{\circ}$ and 50$^{\circ}$,  which again corresponds to the location where the radial magnetic field is minimum. One difference is the larger abundance area extends to a slightly lower latitude than is seen in the case of chromium, seen at latitudes -40$^{\circ}$ to 30$^{\circ}$ and at most longitudes. Comparing to the magnetic field map,  again this is located in areas where the radial field modulus is on the order of 2 kG.

\subsection{Rare Earth Elements - Praseodymium, Neodymium, Europium} 
\begin{landscape}
\begin{figure}
 \includegraphics[width=0.38\textwidth]{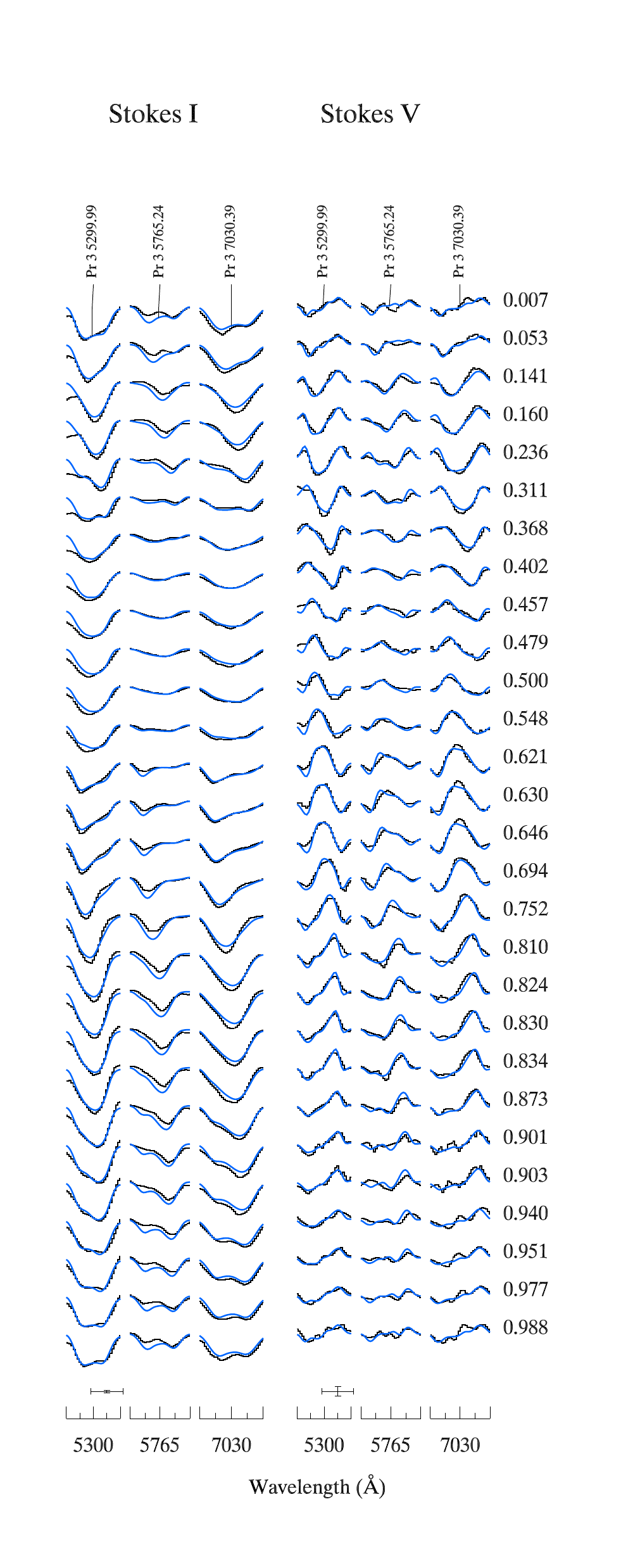}
\includegraphics[width=0.49\textwidth]{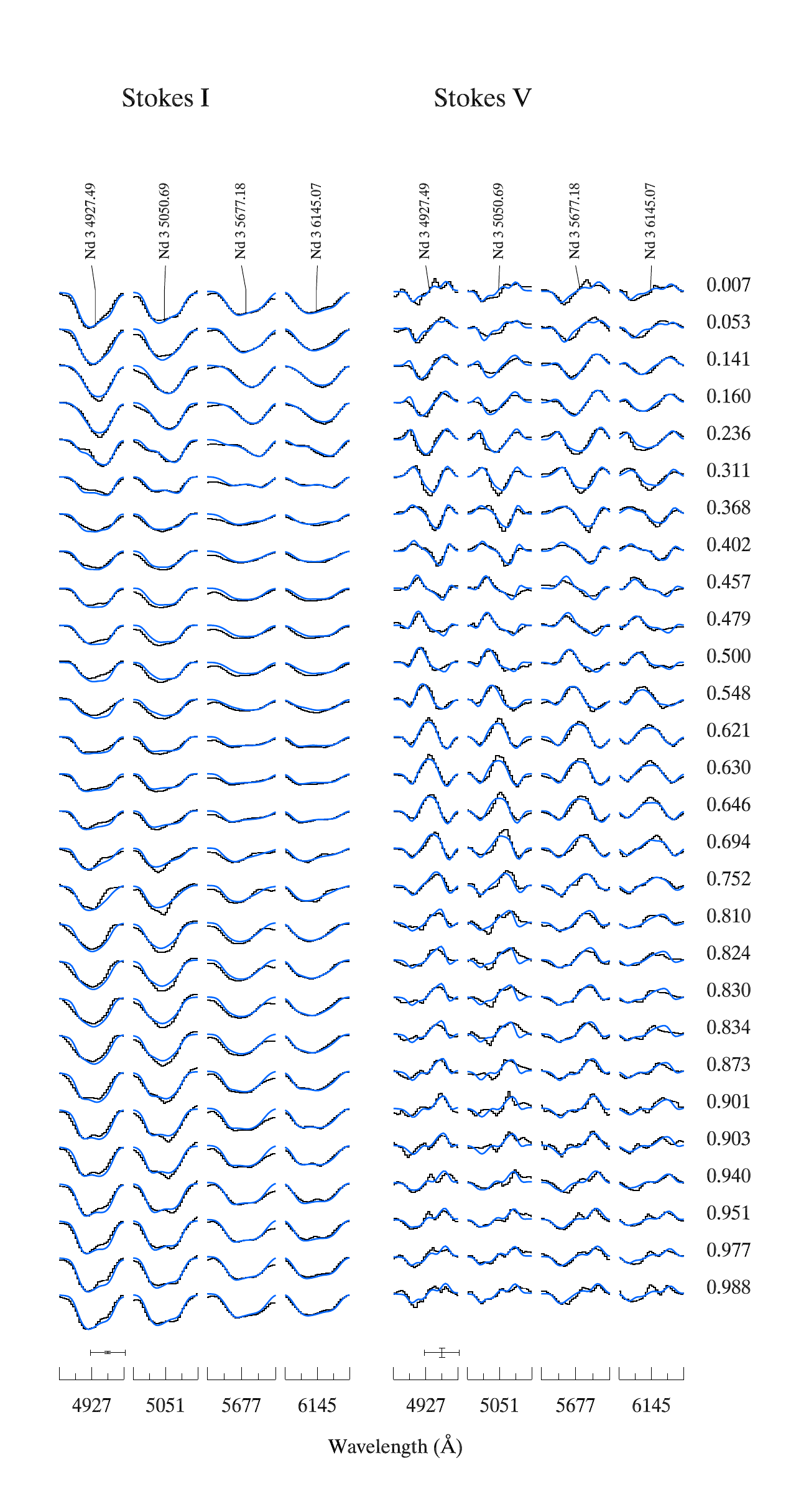}
  \includegraphics[width=0.33\textwidth]{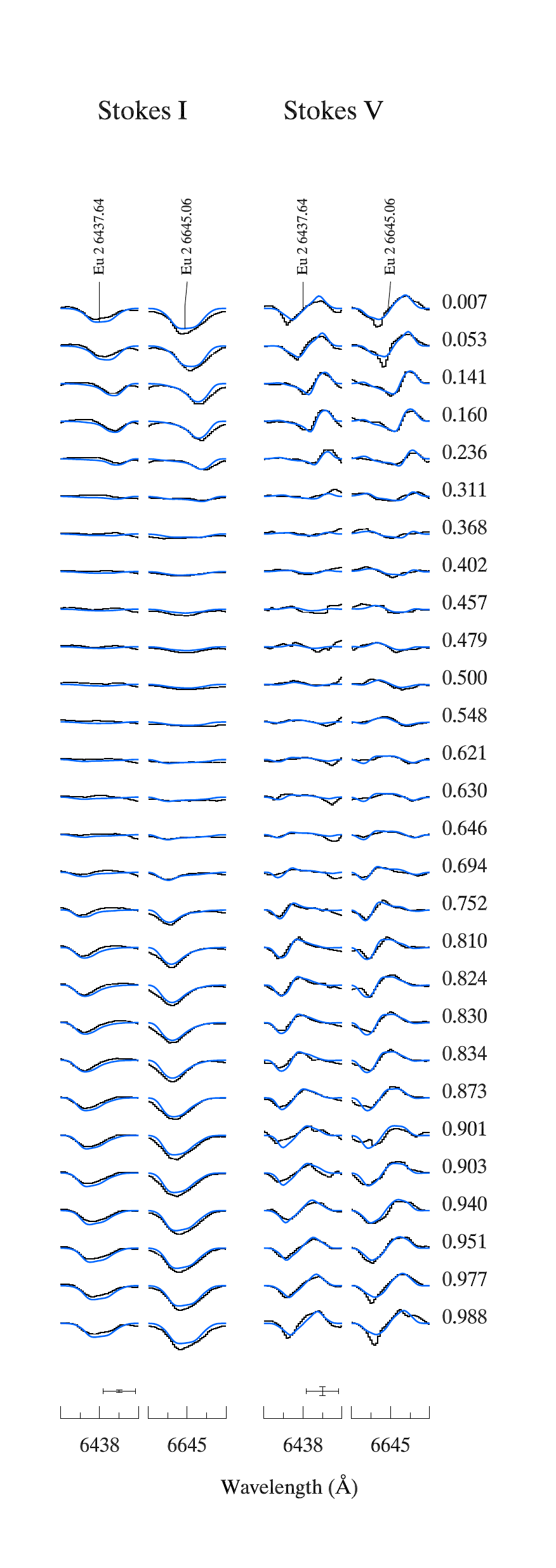}
   \caption{Comparison between observed (dots) and synthetic (solid curve,blue) Stokes $IV$ parameter spectra of $\alpha^2$  using INVERS10. The solid curves shows the best fit to Pr, Nd and Eu lines }
\label{Pr-Nd-Eu-fit-IV}
\end{figure}
\end{landscape}

Praseodymium chemical maps were produced using the Stokes $IV$ line profiles of Pr~{\sc iii}\ $\lambda$ 5299, 5765 and 7030. The fit between observations and the model can be seen in Fig. \ref{Pr-Nd-Eu-fit-IV} and the map is shown in Fig. \ref{Maps2}. In Fig. \ref{Pr-Nd-Eu-fit-IV}, the lines show strong variability in the Stokes $I$ profile,  with a clear change in the line shape between phases.  There is agreement between observations and model at most phases, with the $\lambda$ 5299 wing not fully fit.  The praseodymium abundance ranges from -5.0 dex to -8.0 dex  (solar value = -11.33), the abundance distribution shows two areas of higher abundance at latitudes between -30$^{\circ}$ to +30$^{\circ}$, and at longitudes between 20$^{\circ}$ to 100$^{\circ}$ and 300$^{\circ}$ to 340$^{\circ}$. When comparing to the magnetic field a direct correlation is not clear, but the location of the high abundance areas overlap with areas where the negative radial field is greatest and the depleted areas appear to trace the magnetic equator. 

Neodymium chemical maps were produced using the Stokes $IV$ line profiles of Nd~{\sc iii}\ $\lambda$ 4927, 5050, 5677 and 6145. The fit between observations and the model can be seen in Fig. \ref{Pr-Nd-Eu-fit-IV} and the map is shown in Fig. \ref{Maps2}. In Fig. \ref{Pr-Nd-Eu-fit-IV}, the lines show some variability in the Stokes $I$ profile,  with subtle line shape changes between phases. There is good agreement between observations and model at most phases. The neodymium abundance ranges from -5.0 dex to -9.0 dex  (solar value = -10.54). Similar to titanium, the higher abundance areas are found at latitudes of 15$^{\circ}$ to 60$^{\circ}$ and spread over all latitudes. When comparing this distribution with the magnetic field map,  the higher abundance areas appear to correlate with areas where the radial field modulus is 2 kG, regardless if it is the positive or negative component. As with titanium there are also small discrete low abundance areas seen at longitudes 120$^{\circ}$ and 260$^{\circ}$ extending to just above the stellar equator, areas where the radial field is small, in the transition region between negative and positive components. 

Europium chemical maps were produced using the Stokes $IV$ line profiles of Eu~{\sc ii}\ $\lambda$ 6437 and 6645. The fit between observations and the model can be seen in Fig. \ref{Pr-Nd-Eu-fit-IV} and the map is shown in Fig. \ref{Maps2}. In Fig. \ref{Pr-Nd-Eu-fit-IV}, both lines show strong variability in the Stokes $I$ profile, with generally good agreement between observations and model at most phases. It should be noted that these europium lines have hyperfine and isotopic structure (which is not included in the inversions). An experiment was performed to see if by including these structures in the inversion, an improved fit was found.  The result was that the fit to the Stokes $IV$ profiles was no better than when hyperfine and isotopic structure were ignored and the resulting abundance map was very similar. The europium abundance ranges from -5.5 dex to -8.0 dex  (solar value = -11.53), with the structure very similar to that seen for oxygen. The highest concentrations are found at latitudes between -45$^{\circ}$ to +45$^{\circ}$ and a longitude between 320$^{\circ}$ and 40$^{\circ}$ which aligns with the negative part of the radial magnetic field.  There is an extended low abundance area (seen at longitudes between 100$^{\circ}$ and 300$^{\circ}$) which is seen at the location where the radial magnetic field is negligible and it traces the boundary around the magnetic equator. 

\begin{table*}
\begin{center}
\caption{Summary of the result of the derived abundance maps for $\alpha^2$ CVn}
\begin{tabular}{cccc}
\hline
Element  & Location of enhancement & Location where depleted & General structure \\
\hline
O  &  At negative pole      &  Where field is weak and positive pole  &  Large single spot / structure \\
Si   & Where field $\approx$ 2 kG  &  Where field is weak & Distributed at high latitudes \\
Cl & At negative pole     &  Where field is weak and positive pole  & Large single spot / structure \\
Ti  & Where field $\approx$ 2 kG  &  Where field is weak & Distributed at high latitudes \\
Cr  & Where field $\approx$ 2 kG  &  Where field is weak & Distributed with two depleted spot regions \\
Fe & Where field $\approx$ 2 kG  &  Where field is weak & Distributed with depleted spot region \\
Pr & At negative pole     &  Magnetic equator and positive pole  & Two spots / structures \\
Nd & Where field  $\approx$ 2 kG  &  Where field is weak & Distributed at high latitudes \\
Eu & At negative pole      &  Magnetic equator and positive pole & Large single spot / structure \\
\hline
\label{abundance-table}
\end{tabular}
\end{center}
\end{table*}

\section{Further Tests}
\begin{figure}
\begin{center}
       \includegraphics[width=0.51\textwidth]{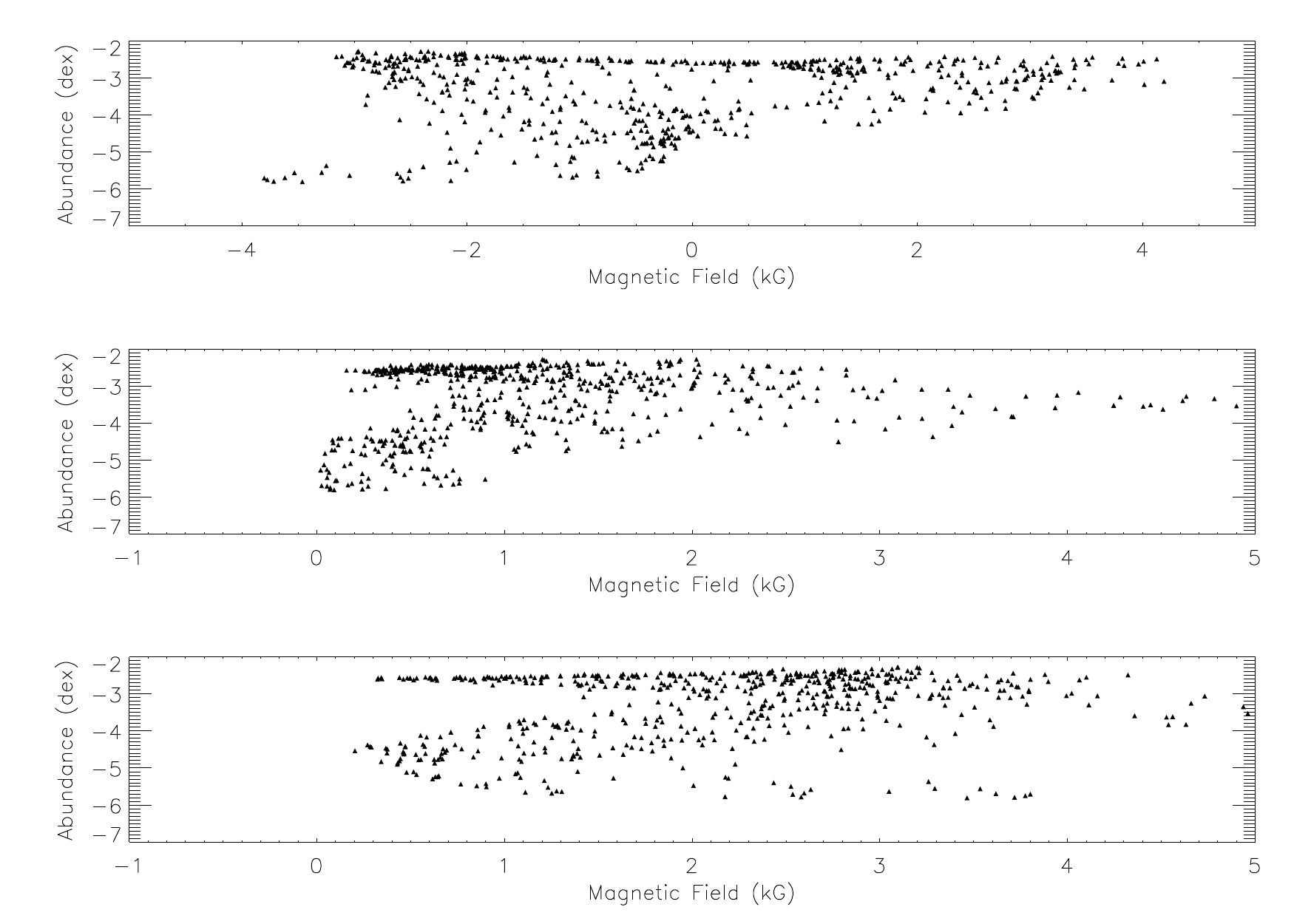}
     \caption{Scatter plots for iron abundance vs the radial magnetic field (top), the horizontal magnetic field (middle) and the magnetic field modulus (bottom). Each point represents one surface element of the reconstructed visible stellar surface.}
\label{scatter}
\end{center}
\end{figure}

In an attempt to more quantitatively investigate potential correlations between the abundance and magnetic field structures we produced a series of scatter plots of abundance values at each surface element (for the visible part of the stellar surface) as a function of different magnetic field properties at the same surface element (the magnetic field modulus, the horizontal magnetic field and the radial magnetic field). The resulting scatter plots however did not reveal any clear correlations and provided no significant information in addition to that already recovered by visual comparison of the abundance structure and magnetic field maps.  An example of such a scatter plot is shown for the iron abundance in Fig \ref{scatter}. 

An additional test was to seek evidence of temporal evolution of abundance structures using Eu~{\sc ii} lines reported by by Farnsworth (1932) and our new observations. Farnsworth utilised equivalent width measurements of Eu~{\sc ii} lines to determine the rotational period of  $\alpha^2$~CVn. We used a selection of the same Eu~{\sc ii} lines (Eu~{\sc ii} $\lambda$ 4132, 4205 and 6645) and measured the equivalent width of these lines at each phase. The equivalent widths were plotted as a function of phase using the rotational ephemeris determined by Farnsworth (1932). We find the maximum equivalent width generally occurs at zero phase within the effective measurement uncertainties, in agreement with Farnsworth(1932). This result indicates that there is no evidence for large-scale Eu spot evolution in  $\alpha^2$~CVn.

Our additional tests also show that there is no evidence for evolution of the detailed surface abundance structure over a decade timescale, and no change int he large-scale distribution of Eu since 1932.

\section{Conclusion and Discussion}
We presented a slightly modified and updated magnetic field map of $\alpha^2$ CVn which was required to improve the fit between the model line profiles and observed line profiles in Stokes $V$ for the elements oxygen and chlorine. We adopted this magnetic map as our fixed field map in all our inversions. This new magnetic map which was reconstructed using a lower value of magnetic field regularisation than used by Silvester et al. (2014), is similar to what is seen by Silvester et al. (2014), with an additional small magnetic spot-like region seen at phase 0.00.  The inclusion of oxygen and chlorine resulted in this new structures seen at phase 0.00. Overall we find that this updated map does not change the findings of Silvester et al. (2014).

Silvester et al. (2014) used the constraint that we only included lines with clear Stokes $QU$ signatures in the inversions, thus lines such as oxygen and chlorine were not considered. It should be noted that the map that derived by Silvester et al. (2014) is not what we would consider the ultimate magnetic field map of $\alpha^2$ CVn, rather the most realistic representation of the magnetic field taking into account the aforementioned constraint. One consideration we did not previously take into account,  was to include lines which exhibited strong abundance structures at opposing phases, where elements sample the stellar surface in a different way. 

The resulting chemical abundance maps can be classified into distinct groups;  those that accumulate close to the negative pole or the negative part of the radial field and those that are located near the stellar equator where the field is close to a field modulus value of around 2 kG.  The results are summarised in Table \ref{abundance-table}.

The elements oxygen, chlorine and europium exhibit one large enhanced abundance structure seen between phases 0.80 to 0.20. This location correlates to the negative radial field.  The praseodymium abundance structure is similar to that of oxygen, chlorine and europium, but with an 0.20 phase offset and with a second enhanced structure. These praseodymium structures appear to only align with the negative radial magnetic field. Unlike the other five elements studied, these elements show enhanced abundances only near the negative radial field and not in the areas of the field which have a strong positive radial component. Furthermore both europium and praseodymium have low abundance areas which trace the magnetic equator. 

Rice et al. (2004) studied the distribution of oxygen on the surface of the Ap star $\theta$ Aur and found that oxygen was lower in abundance around the magnetic equator and enhanced in bands around the magnetic poles. For the Ap star HD 3980 Nesvacil at al. (2012) found that in the case of oxygen there were circular areas of high abundance around both magnetic poles and that the abundance was depleted around the magnetic equator. In the case of the roAp HD 83368 (Kochukhov et al. 2004) they found oxygen enhanced at the magnetic equator. None of these studies found oxygen located only on one magnetic pole. Chlorine has rarely been mapped in an Ap star. Kochukhov at al. (2002) mapped chlorine on $\alpha^2$ CVn and found a very similar distribution for chlorine as we find. For HD 3980 Nesvacil et al. (2012) report that the europium enhancement is located in spots between the magnetic poles and the magnetic equator. In the case of praseodymium they found high abundance regions in the area of the magnetic poles, in our case we find the areas of high abundance are located on the negative radial component and not the positive component of the field.  

Iron and chromium were found to have depleted abundance at areas where the magnetic field is weakest and the higher abundance features trace the areas of the magnetic field where the magnetic field modulus is on the order of 2 kG. This is somewhat consistent with what was found for Cr and Fe in HD 3980 where enhancements were found around the poles and the stellar equator (Nesvacil at al. 2012) and for $\varepsilon$ UMa  (L\"{u}ftinger et al. 2003) where both Cr and Fe were enhanced at the poles. 

In the case of the elements silicon, titanium and neodymium the higher abundance areas are broadly distributed around the region where the field modulus is around 2.0 kG, but unlike chromium and iron the depleted abundance regions are larger and cover the lower half of the stellar sphere at most phases, where the field is consistently weak. Looking to previous studies, we find no clear correlation for these elements. Nesvacil et al. (2012) found for silicon on HD 3980 that areas of overabundance were located at various spots along the rotational equator between the magnetic poles and magnetic equator. In the case of titanium, L\"{u}ftinger et al. (2003) found that titanium in $\varepsilon$ UMa was accumulated around the magnetic equator. Neodymium on HD 3980 was found to be concentrated at the magnetic poles and depleted at the magnetic equator (Nesvacil et al. 2012). 

It is clear from our derived abundance maps that the magnetic field has an influence on the formation of horizontal abundance structure for all elements, but by looking at previous mapping result with the exception of Fe and Cr, there appears no clear correlation between our result and results for other Ap stars. In particular given elements don't correlate to the same magnetic field regions between respective studies. 

Michaud et al. (1981) predicted that in the absence of turbulence, rare earth elements should be concentrated where the magnetic field is horizontal (magnetic equator) and iron should be enhanced where the field is vertical (magnetic poles). This is not clearly seen in our maps. In the case of the elements europium and praseodymium they are depleted at the magnetic equator and not enhanced. Whilst iron and chromium are enhanced near the poles, the region of enhancement is much larger, almost to the magnetic equator in some cases.  For silicon Alecian \& Vauclair (1981) predicted that enhancements will be seen at the magnetic equator. Again this is not seen in our maps, with silicon being distributed over a large area. Alecian \& Stift (2010) comment that if the magnetic field of a star is not purely dipolar, the proposed belt-like enhancements of some elements at the magnetic equator may be too small to be detectable using current Doppler mapping techniques and would result in the appearance of spots in these regions instead.  Such spots are not seen in our maps at the magnetic equator. It is also interesting to note that we don't find any abundance structures for any of the elements studied which we can directly relate to the small-scale structures of the magnetic field modulus.

This all suggests that important details are missing from the theory relating to the formation of horizontal abundance structures and the magnetic field and that a better understanding of the vertical abundance structure and the impact the magnetic field has on this structure, is ultimately required. Our additional tests also show that there is no evidence for abundance structure variability over a decade timescale. To obtain further observational constraints,  it would be useful to map additional Ap stars. This will be investigated in future papers using the Stokes $IQUV$ data of other Ap stars, in part collected by Silvester et al. (2012).  In addition mapping the roAp star HD 24712 in Stokes $IQUV$ (by Rusomarov et al. 2013) is ongoing, which could potentially give additional constraints. 

\section*{Acknowledgments} 
OK is a Royal Swedish Academy of Sciences Research Fellow supported by grants from the Knut and Alice Wallenberg Foundation,  the Swedish Research Council and G\"{o}ran Gustafsson Foundation.
GAW acknowledges support from the Natural Science and Engineering Research Council of Canada in the form of a Discovery Grant.
JS thanks Dr Kristine Spekkens for providing a workstation on which to run {\sc invers}10.

\bsp

\label{lastpage}

\end{document}